\documentclass[aps,prd,preprintnumbers,superscriptaddress,nofootinbib,onecolumn,notitlepage]{revtex4-1}
\usepackage{ulem}
\usepackage[dvipdfmx]{graphicx}
\usepackage{bm,latexsym,amsmath,amssymb,amsfonts}
\usepackage[colorlinks=true,linkcolor=magenta,citecolor=magenta]{hyperref}
\usepackage{color}
\usepackage{framed}

\begin{document}

\title{
Resolving a spacetime singularity with field transformations
}

\author{Atsushi Naruko}
\affiliation{Frontier Research Institute for Interdisciplinary Sciences \& Department of Physics, \\
Tohoku University, Sendai 980-8578, Japan}
\affiliation{Center for Gravitational Physics, Yukawa Institute for Theoretical Physics, Kyoto University, Kyoto 606-8502, Japan} 

\author{Chul-Moon Yoo}
\affiliation{Division of Particle and Astrophysical Science, Graduate School of Science, \\
Nagoya University, Nagoya 464-8602, Japan}

\author{Misao Sasaki}
\affiliation{Center for Gravitational Physics, Yukawa Institute for Theoretical Physics, Kyoto University, Kyoto 606-8502, Japan} 
\affiliation{Kavli Institute for the Physics and Mathematics of the Universe (WPI), The University of Tokyo, Chiba 277-8583, Japan}
\affiliation{Leung Center for Cosmology and Particle Astrophysics, National Taiwan University, Taipei 10617, Taiwan}

\preprint{TU1083,YITP-19-03}

\begin{abstract}
It is widely believed that classical gravity breaks down and quantum gravity is needed to deal with a singularity.
We show that there is a class of spacetime curvature singularities which can be resolved with metric and matter field transformations.
As an example, we consider an anisotropic power-law inflation model with scalar and gauge fields in which a space-like curvature singularity exists at the beginning of time. 
First, we provide a transformation of the metric to the flat geometry, i.e. the Minkowski metric. 
The transformation removes the curvature singularity located at the origin of the time.
An essential difference from previous work in the literature is that the origin of time is not sent to past infinity by the transformation but it remains at a finite time in the past. 
Thus the geometry becomes extendible beyond the singularity.
In general, matter fields are still singular in their original form after such a metric transformation. 
However, we explicitly show that there is a case in which the singular behavior of the matter fields can be completely removed by a re-definition of matter fields.
Thus, for the first time, we have resolved a class of initial cosmic singularities and successfully extended the spacetime beyond the singularity in the framework of classical gravity. 
\end{abstract}

\maketitle

\section{Introduction} 
\label{sec1}

A spacetime singularity often appears as a boundary of spacetime where
 the spacetime curvature is divergent.
In the vicinity of a spacetime singularity,
 all the known fundamental laws of physics lose their predictabilities and become useless. 
It would be natural to expect that a resolution of a spacetime singularity would be completed by
 a more fundamental theory, such as a quantum theory of gravity. 
Thus, a spacetime singularity can be a testing site for new ideas in the frontier of the theoretical physics. 

Our aim in this paper is not to propose a way to resolve a spacetime singularity
 with a new fundamental theory, but just to point out a possible different aspect
 for a class of spacetime singularities. 
The key idea is the use of metric transformations, 
namely conformal and disformal metric transformations~\cite{Bekenstein:1992pj}. 
Recently, metric transformations of a theory of gravity with a scalar field have been extensively investigated~\cite{Deruelle:2010ht,Bettoni:2013diz,Chiba:2013mha,Zumalacarregui:2013pma,Deruelle:2014zza,Domenech:2015qoa,Domenech:2015hka,Arroja:2015wpa,Domenech:2015tca,Crisostomi:2016czh,Achour:2016rkg} and those with a vector field~\cite{%Jimenez:2016isa,
Heisenberg:2016eld,Heisenberg:2014rta,Kimura:2016rzw,Papadopoulos:2017xxx,Domenech:2018vqj}. 
It is known that, as long as a metric transformation is regular and invertible,
the nature of the theory, such as the number of degrees of freedom,
will be preserved. 

Let us consider a solution of a set of spacetime and matter fields in
a particular theory in which a singularity exists.
By performing a metric transformation which is regular,
we obtain an equivalent theory of gravity with matter fields.
If the transformation is singular at the singularity,
 the resultant structure of the spacetime may be altered. 
For example, it is possible to create a real curvature singularity with singular matter field from completely regular spacetime with regular matter by means of a singular metric transformation. 
Conversely, we may be able to eliminate some class of spacetime singularities with singular matter fields by a singular field transformation to obtain a perfectly regular spacetime with regular matter fields.

We note that a conformal and disformal metric transformation considered in this
paper is not completely arbitrary but must be described by a matter field involved
in the original theory so that one can obtain an equivalent theory, except at the spacetime singularity.
For instance, a singularity of the Schwarzschild solution cannot be eliminated by our procedure 
 since it is a vacuum solution without matter field. 
In this paper, we will particularly focus on a class of cosmological solutions with matter fields
 where a space-like singularity exists.

This paper is organized as follows. 
We introduce an anisotropic power-law inflation model, as an example,
 in which a space-like cosmological singularity exists at the origin of time.
We demonstrate that the metric of the anisotropic power-law inflation model can be transformed into
 a flat metric after a combination of conformal and disformal metric transformations.
Although the matter fields remain singular in their original form, 
we show that they may be also regularized by a field redefinition of matter fields,
provided that the action is regular. Thus the regularity of the action is
essential for the complete removal of a spacetime singularity.

\section{Anisotropic universe model}
\label{sec2}
We first discuss the impact of a metric transformation on a spacetime singularity.
To make discussion concrete, we invoke an explicit model dubbed  
 the anisotropic power-law inflation~\cite{Kanno:2010nr}. 
Hereafter, we adopt the Planck units $8\pi G=M_\mathrm{pl}^2 = 1$.

Let us consider a scalar and a $U (1)$ gauge field coupled with Einstein gravity whose action is given by
 \begin{align}
 S = \int {\rm d}^4 x \sqrt{- g} \left[ \frac{1}{2 } R - \frac{1}{2} (\nabla \phi)^2 - V_0 e^{\lambda \phi}
 - \frac{1}{4} \, f_0^2 \, e^{2 \rho \phi} \, F_{\mu \nu} F^{\mu \nu} \right] \,. 
 \label{action:aniso}
 \end{align}
Here, we are interested in a homogeneous but anisotropic background spacetime
 and hence the vector field will be present at the background and will be assumed to be homogeneous.
We fix the $U(1)$ gauge by setting $A_0=0$. 
The spatial direction of the vector field can be identified with the $x-$direction. 
Thus, we assume the vector to be $A_\mu = \bigl( 0, A (t), 0, 0 \bigr)$.
As for spacetime, we assume the Bianchi type-I metric with constant $\alpha$ and $\beta$:
 \begin{align}
 {\rm d} s^2 = - {\rm d} t^2 + t^{2 \alpha} \Bigl[ t^{-4 \beta} {\rm d} x^2 + t^{2 \beta} ({\rm d} y^2 + {\rm d} z^2) \Bigr], 
 \end{align}
where we have assumed the rotational symmetry on the $y$--$z$ plane and 
the power-law expansion in each spatial direction. 

From the matter and gravitational field equations, it is not difficult to find 
a particular solution (see Ref.~\cite{Kanno:2010nr} for a complete expression): 
\begin{subequations}
 \begin{align}
 \phi(t) &= - \frac{2}{\lambda} \log t \,, \\
 A(t) &=t^{-1 + 3 \alpha} \,, \\
 \alpha &= \frac{4}{3 \lambda (\lambda+2 \rho) }+\frac{\rho }{\lambda }+\frac{1}{6} \,, 
 \label{alpha}
 \\
 \beta &= \frac{1}{3}-\frac{4}{3 \lambda (\lambda + 2 \rho) } \,, 
\label{beta}
 \end{align}
 \label{sol}
\end{subequations}
where $\lambda$ and $\rho$ are parameters that characterize $f_0^2$ and $V_0$.
Hereafter we regard $\alpha$ and $\beta$ as independent parameters which characterize
 the solutions since $\rho$ and $\lambda$ can be also expressed as functions of them inversely. 

It is important to know that there is a curvature singularity at the origin of time, $t=0$.
It can be checked by looking at, for example,
 the Kretchmann invariant scalar $I_R\equiv R_{\mu\nu\rho\sigma}R^{\mu\nu\rho\sigma}$ and 
 the Weyl scalar $I_W\equiv W_{\mu\nu\rho\sigma}W^{\mu\nu\rho\sigma}$ 
 where $R_{\mu\nu\rho\sigma}$ and $W_{\mu\nu\rho\sigma}$ are the Riemann and 
 Weyl curvature tensors, respectively.
For the above metric, the explicit form of these invariants are given by
\begin{align}
I_R &= \frac{12}{t^4} \Bigl[ (2 \alpha ^2-2 \alpha +1) \alpha^2 + 2 (6 \alpha ^2-6 \alpha +1) \beta ^2 
 - 4 (\alpha -1) \beta^3 + 9 \beta^4 \Bigr] \,, \\
I_W &= \frac{12}{t^4} (1 - \alpha + 2 \beta)^2 \beta^2 \,.
\end{align}
It is apparent that these two quantities are divergent at $t=0$ in general, 
which implies the existence of a space-like singularity at the origin of time. 

We should note that the Weyl scalar is invariant under a conformal transformation
 and hence its divergence cannot be cured by a conformal transformation. 
 On the other hand, it is not invariant under a disformal transformation. 
 In the following sections, we explicitly show that the divergence of the Weyl curvature 
 can be also resolved by utilising a disformal transformation associated with the vector field 
 \footnote{We also note that we can consider a simpler example of a homogeneous and isotropic universe 
where the Weyl curvature vanishes.
 In this case, the singularity can be resolved by using only a conformal transformation.}.

\section{Metric transformation to a flat metric}
\label{sec3}
In the anisotropic universe model, we have
 $A_\mu {\rm d} x^\mu = A_x (t) \, {\rm d} x = A (t) \, {\rm d} x$. 
We can rewrite the metric by introducing the vector field:
 \begin{align}
 {\rm d} s^2 
 &= (\gamma \, \tau)^{2 (\alpha + \beta)/\gamma} \biggl[ - {\rm d} \tau^2 + {\rm d} x^2 + {\rm d} y^2 + {\rm d} z^2 
  + \bigl[ (\gamma \, \tau)^{- 6 \beta/\gamma} - 1 \bigr] \left( \frac{A_\mu {\rm d} x^\mu}{A (\tau)} \right)^2 
  \biggr] \,.
  \label{soda-metric}
 \end{align}
Here, we have introduced the conformal time 
 by ${\rm d} \tau = t^{\gamma-1} {\rm d} t$ and hence $\gamma \tau = t^\gamma$ with  $\gamma = 1 -(\alpha +\beta)$.
Since all the quantities can be expressed in terms of $\tau (t)$ in the current set-up, 
we can also regard them as functions of the scalar field and the contraction of the vector field, which is defined by $Y \equiv \eta^{\mu\nu} A_\mu A_\nu 
=(\gamma \tau)^{2 (-1+3 \alpha)/\gamma}$.
We note that not the original metric but the flat one was used to define
the contraction of the vector.
Remembering that $t = (\gamma \, \tau)^{1/\gamma} = \exp[-(\lambda \phi)/2] = Y^\delta$,
where  $2\delta \equiv (-1 + 3 \alpha)^{-1}$,
it is not difficult to re-express (\ref{soda-metric}) as
 \begin{subequations}
 \begin{align}
 {\rm d} s^2 
 &= \Omega^2 (\phi) \Bigl[ \eta_{\mu \nu} + \Gamma (Y) A_\mu A_\nu \Bigr] {\rm d} x^\mu {\rm d} x^\nu \,, 
  \end{align}
 where
\begin{align}
 \Omega (\phi) 
 &= (\gamma \, \tau)^{2 (\alpha + \beta)/\gamma} 
 = \exp[ - (\alpha + \beta) \lambda \phi] \,, 
  \label{sol:omega} \\ 
 \Gamma (Y) 
 &= \bigl[ (\gamma \, \tau)^{- 6 \beta/\gamma} - 1 \bigr]/A^2
 = Y^{-1}  (Y^{- 6 \beta \delta} - 1) \,.
 \label{sol:gamma}
 \end{align}
 \end{subequations}
Now it is trivial to see that the metric can be converted to a flat one by 
conformal and disformal metric transformations with the conformal and 
disformal factors given by Eqs.~(\ref{sol:omega}) and (\ref{sol:gamma}).

It should be noted that the effective stress-energy tensor described by the scalar and gauge fields in the resultant theory vanish since the Einstein tensor trivially vanishes in the Minkowski spacetime.
Thus, the scalar and gauge fields are stealth fields and do not contribute to the effective 
energy momentum tensor even though the field values are divergent at the singularity.  
In other words, at this stage we just hide the spacetime singularity behind the singular stealth matter 
fields.\footnote{One can show that the time-like singularity in the 
	Reissner-Nordstr\"om spacetime can be also hidden by a similar transformation
	 to Eq.~\eqref{soda-metric}.
}
Because of the singular behavior of the matter fields,
 the theory may not be well-defined yet at the singularity. 
Nevertheless, in the following section, we explicitly show that there is a specific case where 
the singular behavior of the matter field can be regularized by performing a re-definition of matter fields.

\section{Regularity of the action and matter fields}
\label{sec4}
We show that there is a class of singularities
which can be totally resolved and the resultant theory after the transformation is 
completely well-defined even beyond the point corresponding to the original singularity. 

As we have seen above, there is a space-like singularity at $t=0$ in the solution discussed in
the previous section. While we successfully removed the spacetime singularity from the geometry,  
the matter fields remain singular in general.
At this point, it is important to note the regularity of the action.
Since the action is invariant under any transformation of the field variables including
the metric, it would be impossible to find any transformation that could regularize a singularity if the original action is singular.
Conversely, if the action is regular, we may be able to find a set of conformal and
disformal metric transformations as well as matter field transformations that can completely remove the singularity in the matter fields as well as the metric.
Therefore, below we focus on the case where the regularity of the action is guaranteed,
and look for transformations that make the metric and matter fields regular simultaneously. 

By plugging the solutions in \eqref{sol}, into the action in Eq (\ref{action:aniso}), one reads
\begin{align}
 S 
&= \int {\rm d} \tau \, {\rm d}^3 x \, t^{4 \alpha + \beta - 2} (-1+ 3\alpha) (\alpha + \beta) 
= \int {\rm d} \tau \, {\rm d}^3 x \, (\gamma \tau)^{(4 \alpha + \beta - 2)/\gamma} (-1+ 3\alpha) (\alpha + \beta) \,. 
\end{align}
It is clear that the action is singular at the singularity ($t=\tau = 0$) if $4 \alpha + \beta - 2< 0$ and $\gamma > 0$.
In order to make the action regular at the singularity, we consider a specific case with $\alpha = 2/3$ and $\beta = 4/21$, corresponding to $\lambda=7/3$ and $\rho = 5/6$. 
With this choice of parameters, the action reduces to 
\begin{align}
 S 
= \frac{6}{7^7} \int {\rm d} \tau \, {\rm d}^3 x \, \tau^6 \,. 
 \label{action}
\end{align}
whose regularity at $t=\tau=0$ is manifest.
Thanks to the special choice of parameters, the gauge field vanishes at the singularity and also its time derivative is always regular since $A (t) = t$ and $\dot{A} (t) = 1$.
On the other hand, the scalar field remains singular at the singularity since
\begin{align}
 \phi = - \frac{2}{\lambda} \log t 
 = - \frac{2}{\lambda \gamma} \log (\gamma \tau)
 = - 6 \log(\tau/7) \,.
\label{action:phi}
\end{align}
Hence, the Lagrangian of the scalar field is singular too since $\dot{\phi}^2 \propto t^{-2}\propto \tau^{-14}$. 
However, fortunately, the metric determinant factor gives $\sqrt{-g} \propto t^{4 \alpha} = t^{8/3}\propto \tau^{56/3}$.
Thus the Lagrangian density is regular even at the origin of time. 
This implies if we perform conformal and disformal metric transformations to
make the metric flat, we will obtain a regular Lagrangian.

Denoting the metric in the original frame by $\widetilde{g}_{\mu\nu}$, we perform the conformal and disformal metric transformation,
 $\widetilde{g}_{\mu \nu} = \Omega^2 (g_{\mu \nu} + \Gamma A_\mu A_\nu)$ with $\Omega=e^{-2\phi}$.
Then the action (\ref{action:aniso}) reduces to
 \begin{align}
 S 
 &= \int {\rm d}^4 x \sqrt{-\widetilde{g}}
  \left[ \frac{1}{2} \widetilde{R} - \frac{1}{2} (\widetilde{\nabla} \phi)^2 - V (\phi) 
 - \frac{1}{4} f^2 \widetilde{g}^{\mu \rho} \widetilde{g}^{\nu \sigma} F_{\mu \nu} F_{\rho \sigma} \right] \notag\\
 &= \int {\rm d}^4 x \, \sqrt{-g} \, \Omega^2 \, \sqrt{1 + Y \Gamma} \biggl\{ \frac{1}{2} R
 - \frac{1}{2} \frac{\Gamma}{1 + Y \Gamma} \Bigl[ (\nabla_\mu A^\mu)^2 
 - (\nabla_\mu A_\nu) (\nabla^\mu A^\nu) \Bigr] \notag\\
 & \qquad 
 + \frac{1}{2} \frac{\Gamma_Y}{1+ Y \Gamma} 
 \Bigl[ (A^\mu S_{\mu \rho}) (A^\nu S_{\nu}{}^\rho) 
 - 2 (A^\mu A^\nu \nabla_{\mu} A_\nu) (\nabla_\rho A^\rho) \Bigr] \notag\\
 & \qquad 
 - \frac{1}{8} \Gamma (F_{\mu \nu})^2
 + \frac{1}{4} \frac{\Gamma^2 - \Gamma_Y}{1 + Y \Gamma} (A^\mu F_{\mu \rho}) (A^\nu F_{\nu}{}^\rho) \notag\\
 & \qquad 
 - \frac{3}{\Omega \sqrt{1 + Y \Gamma} \sqrt{- g}} \partial_\mu \left[ \sqrt{- g} \sqrt{1 + Y \Gamma} \left( g^{\mu \nu} - \frac{\Gamma}{1 + Y \Gamma} A^\mu A^\nu \right) \partial_\nu \Omega \right] \notag\\
 & \qquad 
 - \frac{1}{8} \frac{1}{\Omega^2} \left( g^{\mu \nu} - \frac{\Gamma}{1 + Y \Gamma} A^\mu A^\nu \right) 
 (\partial_\mu \Omega) (\partial_\nu \Omega) - V \notag\\
 & \qquad  
 - \frac{1}{\Omega^2} \frac{1}{4} f^2 \left( g^{\mu \nu} - \frac{\Gamma}{1 + Y \Gamma} A^\mu A^\nu \right)
 \left( g^{\rho \sigma} - \frac{\Gamma}{1 + Y \Gamma} A^\rho A^\sigma \right) F_{\mu \nu} F_{\rho \sigma} 
 \biggr\} \,,
\label{action:new}
\end{align}
where the conformal factor $\Omega$ itself may be regarded as a redefinition of
the original scalar field $\phi$.
Here the explicit form of functions appeared in \eqref{action:new} is given by
\begin{align}
 V &= V_0 e^{\lambda \phi} = \frac{4}{7} e^{(7/3) \phi} = \frac{4}{7} \Omega^{- 7/6} \,, \\
 f &= f_0 e^{\rho \phi} = \frac{2}{\sqrt{7}} e^{(5/6) \phi} = \frac{2}{\sqrt{7}} \Omega^{- 5/12} \,, \\
 \Gamma &= Y^{-1} (Y^{- 6 \beta \delta} - 1) = Y^{-1} (Y^{-4/7} - 1) \,.
\end{align}
For the brevity, we will not explicitly write down the field equations in the new frame 
but the solutions in the new frame will be given by 
\begin{align}
 g_{\mu\nu}=\eta_{\mu\nu} \,, \qquad
\Omega=
%e^{-2 \phi} = 
t^{12/7} = (\tau/7)^{12} \,, \qquad
A = t = (\tau/7)^7 \,.
\label{sol:new}
\end{align}
Thus both the metric and the matter fields $\Omega$ and $A_\mu$ are apparently regular at $t=\tau=0$. 

In conclusion, in the case of  $\alpha = 2/3$ and $\beta = 4/21$ where the value of the action is regular \eqref{action}, we have obtained a manifestly regular solution
that completely removes the cosmological singularity in the original frame at $\tau=0$,
and the solution can be straightforwardly extended to $\tau<0$.
The key observation is the 
the invariance of the action under field transformations and the regularity of the action. 
As we have mentioned earlier, the action is invariant under any transformation of the field variables such as metric as well as matter fields. 
Hence even if original fields are singular at a singularity, 
there may be a set of new variables, which is totally regular and can resolve the singularity, 
as long as the value of the action is regular.

\section{Summary and Discussion}
\label{sec5}
In this paper, we have discussed a possible way to resolve a spacetime singularity 
by transformations of metric and matter fields.
First, we have shown that a certain class of spacetime singularities can be hidden by performing appropriate conformal and disformal metric transformations that are determined by the matter fields involved in the original theory. 
As an explicit example, we have a model of homogeneously but anisotropically expanding universe with a space-like initial singularity, and obtained a set of conformal and disformal metric transformations that make the metric flat.
\footnote{Although we do not present in this paper, it is not difficult to show that
 the time-like singularity in the Reissner-Nordstr\"om spacetime can be also hidden by a similar transformation.} 

The matter fields and associated transformations are singular at the original spacetime singularity, 
and therefore they are also singular in their original form
at the corresponding points in the resultant Minkowski spacetime. 
However, as is trivially realized from the flat geometry, the matter fields do not contribute to 
the effective energy momentum tensor since the Einstein tensor identically vanishes.
Namely, the spacetime singularity is hidden in the singular stealth fields in the resultant theory. 
Then we have shown that for a class of singularities for which the action is regular,
 the singularity associated with the singular stealth fields can be completely eliminated
 by performing a field redefinition of matter fields. 
It is worthy to emphasize that the action is invariant under any transformation of the field variables and hence the regularity of the action is preserved under any transformation even if a transformation is singular.

There are, however, still some open questions, and the significance of our approach 
is not yet clear. For instance, since we have considered only highly symmetric background solutions,
we do not know if perturbations around this background remain regular or not.
Another issue is the change of the spacetime causality through the metric transformation.  
Even though the resultant theory after the transformation is equivalent to the original one
 except for the singularity, interpretations of physical laws in the new theory should be
 significantly different from the original one especially from the point of spacetime causality. 
We should also mention that our approach is apparently not applicable to all 
types of spacetime singularities. 
Therefore the classification of singularities along with similar directions to our work 
may be a not only necessary but also interesting issue.

Despite these issues, our approach to hide/resolve spacetime singularities with field transformations may reveal
 new interesting insights into spacetime singularities, and it could give a hint to overcome 
singularities in a class of gravitational theories without invoking a theory of quantum gravity.

\acknowledgments
We would like to thank Guillem Dom\`enech for useful comments and fruitful discussions. 
A.N. would like to thank Stefano Liberati for an illuminating discussion.
We also would like to thank the Yukawa Institute for Theoretical Physics at Kyoto University, where discussions during the YITP symposium YKIS2018a ``General Relativity -- The Next Generation --" were useful to complete this work.
A.N. is supported in part by a JST grant ``Establishing a Consortium for the Development of Human Resources in Science and Technology''. This work was also partly supported by the JSPS Grant-in-Aid for Scientific Research No.16H01092(A.N.), JP16K17688 and JP19H01895(C.Y.).
This work is supported in part by MEXT grant Nos. 15H05888 and 15K21733.

\end{document}